\documentclass[twocolumn,11pt]{article}
\usepackage[utf8]{inputenc}
\usepackage[T1]{fontenc}
\usepackage{amsmath,amssymb,amsfonts}
\usepackage{graphicx}
\usepackage{algorithm}
\usepackage{algorithmic}
\usepackage{booktabs}
\usepackage{xcolor}
\usepackage{hyperref}
\usepackage{subcaption}
\usepackage{tikz}
\usetikzlibrary{petri,arrows,positioning}
\usepackage[margin=2cm,columnsep=0.5cm]{geometry}
\usepackage{balance}  
\usepackage{pifont}  
\usepackage{enumitem}  
\newcommand{\cmark}{\ding{51}}  

\setlist[itemize]{leftmargin=*,nosep}  
\setlist[enumerate]{leftmargin=*,nosep}

\usepackage{titlesec}
\titlespacing*{\section}{0pt}{8pt plus 2pt minus 2pt}{4pt plus 2pt minus 2pt}
\titlespacing*{\subsection}{0pt}{6pt plus 2pt minus 2pt}{3pt plus 2pt minus 2pt}
\titlespacing*{\subsubsection}{0pt}{4pt plus 2pt minus 2pt}{2pt plus 2pt minus 2pt}

\newcommand{\preset}[1]{\ensuremath{{}^{\bullet}#1}}
\newcommand{\postset}[1]{\ensuremath{#1^{\bullet}}}
\newcommand{\RegSet}[1]{\ensuremath{\Sigma(#1)}}
\newcommand{\DepType}[2]{\ensuremath{\Delta(#1,#2)}}

\title{\textbf{Weak Independence and Coupled Parallelism in Biological Petri Nets}}

\author{
Eugenio Simao\textsuperscript{1}\\
\textsuperscript{1}Universidade Federal de Santa Catarina, UFSC - Ararangua - Brazil\\
eugenio.simao@ufsc.br
}

\date{}

\begin{document}

\maketitle

\begin{abstract}
\textbf{Motivation:} Biological Petri Nets (Bio-PNs) model biochemical pathways where multiple reactions simultaneously affect shared metabolites through convergent production or regulatory coupling. However, classical Petri net independence theory requires transitions to share no places---a constraint that fails to capture biological reality. This mismatch prevents parallel simulation and incorrectly flags biologically valid models as structurally problematic.

\textbf{Results:} To resolve this fundamental limitation, we introduce \emph{weak independence}---a novel formalization distinguishing resource conflicts from biological coupling. Building on this theory, we extend the Bio-PN definition from a classical 5-tuple to a 12-tuple by adding regulatory structure ($\Sigma$), environmental exchange classification ($\Theta$), dependency taxonomy ($\Delta$), heterogeneous transition types ($\tau$), and biochemical formula tracking ($\rho$). This extended formalism enables systematic classification of three place-sharing modes: competitive (conflict), convergent (superposition), and regulatory (read-only). Validating our approach on 100 diverse BioModels (1,775 species, 2,234 reactions across metabolism, signaling, and gene regulation), we find that 96.93\% of transition pairs exhibit weak independence---confirming that biological networks inherently favor cooperation over competition. Our SHYpn implementation demonstrates the practical impact, achieving up to 2.6$\times$ speedup on 30\% of evaluated models.

\textbf{Availability and Implementation:} Open-source at \url{https://github.com/simao-eugenio/shypn} (MIT License).
\end{abstract}

\section{Introduction}

\textbf{Metabolic convergence is ubiquitous in biological systems}. Central metabolism exemplifies this pervasive pattern: glycogenolysis, gluconeogenesis, and starch degradation all converge to produce glucose, while glycolysis and pentose phosphate pathway simultaneously consume it. Beyond metabolism, multi-enzyme complexes catalyze dozens of reactions through shared active sites, and gene regulatory networks exhibit feedback loops where transcription factors regulate multiple genes while themselves being gene products. This concurrent operation is not accidental---it enables homeostasis, rapid environmental response, and efficient resource utilization.

Capturing this biological reality in formal models has proven challenging. Since Reddy et al.'s pioneering work~\cite{Reddy1993}, Petri nets have been successfully applied to biochemical pathways, with extensions for metabolic regulation~\cite{Simao2005} and genetic networks~\cite{Chaouiya2006}. Yet Biological Petri Nets (Bio-PNs) face a fundamental mismatch with classical theory in how they handle \emph{shared places}. Where biological systems routinely exhibit three distinct modes of place-sharing---(1) \textbf{convergent production} (multiple pathways producing the same metabolite), (2) \textbf{regulatory coupling} (reactions sharing enzyme catalysts), and (3) \textbf{competitive consumption} (enzymes competing for substrate)---classical Petri net independence theory treats \emph{all} place-sharing uniformly as potential conflict~\cite{Reisig2013}.

This conservative stance has practical consequences. By preventing parallel execution of any transitions sharing places, classical theory blocks optimization opportunities and incorrectly flags biologically valid models as problematic. The issue becomes critical at scale: BioModels contains over 1,000 curated SBML models (typical: 50--200 species, 100--300 reactions), while genome-scale reconstructions exceed 2,000 reactions~\cite{Orth2011}. For these systems, parameter estimation requiring thousands of simulation runs can take hours sequentially. Meanwhile, the biological reality suggests a different picture---in continuous Bio-PNs with ODE semantics, transitions \emph{can} fire simultaneously when they don't compete for inputs, with rate contributions simply superposing. The formalism thus contradicts the inherent parallelism of molecular interactions.

\subsection{Motivating Example}

\begin{figure}[htbp]
\centering
\begin{tikzpicture}[node distance=1.5cm,>=stealth',bend angle=45,auto]
\tikzstyle{place}=[circle,thick,draw=blue!75,fill=blue!20,minimum size=6mm]
\tikzstyle{transition}=[rectangle,thick,draw=black!75,fill=black!20,minimum size=4mm]

\node [place] (glycogen) {Glycogen};
\node [place, below left=of glycogen] (glucose) {Glucose};
\node [place, below right=of glycogen] (lactate) {Lactate};

\node [transition, left=of glucose] (t1) {$T_1$};
\node [transition, right=of glucose] (t2) {$T_2$};

\draw [->] (glycogen) to node {1} (t1);
\draw [->] (t1) to node {1} (glucose);
\draw [->] (lactate) to node {1} (t2);
\draw [->] (t2) to node {1} (glucose);
\end{tikzpicture}
\caption{Glucose production from two pathways ($T_1$: Glycogenolysis, $T_2$: Gluconeogenesis). Both transitions produce to the same place but don't compete for inputs.}
\label{fig:glucose_example}
\end{figure}
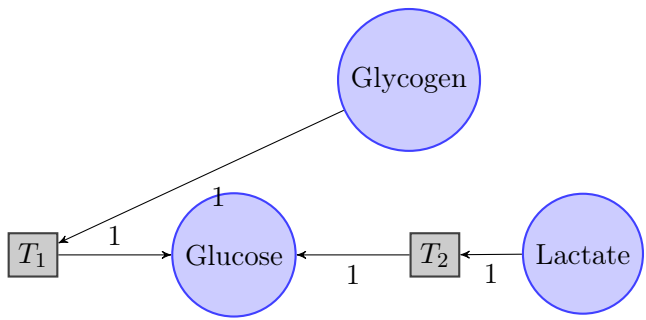

To illustrate the gap concretely, consider the glucose metabolism depicted in Figure~\ref{fig:glucose_example}. Classical analysis declares transitions $T_1$ (glycogenolysis) and $T_2$ (gluconeogenesis) as non-independent because they share output place \texttt{Glucose}. Yet biological reality contradicts this restriction: both pathways can---and do---operate simultaneously, with their production rates combining through mass balance superposition:
\begin{equation}
\frac{d[\text{Glucose}]}{dt} = r_1 + r_2
\end{equation}

\noindent This example captures \textbf{convergent coupling}---a pervasive biological pattern that classical independence theory fails to accommodate. Our weak independence framework recognizes this distinction, enabling faithful representation of concurrent metabolic operation.

\subsection{Contributions}
Addressing these limitations, we present a comprehensive reformulation of Bio-PN foundations spanning theory, formalism, and implementation: (1) \textbf{Weak Independence Theory} formalizes a two-tier independence hierarchy (strong vs.\ weak), enabling parallel execution despite place-sharing when biological semantics permit; (2) \textbf{Extended 12-tuple Bio-PN Definition} systematically incorporates regulatory structure ($\Sigma$), environmental exchange modes ($\Theta$), dependency classification ($\Delta$), heterogeneous transition types ($\tau$), and biochemical formula tracking ($\rho$), unifying continuous, stochastic, and timed dynamics with atomic mass balance validation; (3) \textbf{Dependency Classification Algorithm} ($O(|T|^2 \cdot |P|)$ complexity) distinguishes competitive conflicts from cooperative coupling through systematic analysis; (4) \textbf{Biological Topology Analyzers} provide domain-specific validation (mass balance, flux feasibility) that avoid spurious failures from generic structural checks.

\section{Background and Related Work}

\textbf{Petri Nets in Systems Biology.} Petri nets entered systems biology through Reddy's pioneering work on metabolic pathway modeling (1993)~\cite{Reddy1993}, which established the foundational mapping: places as metabolites, transitions as enzymatic reactions, and arc weights as stoichiometric coefficients. This was followed by Hofestädt's binary PN models for metabolic networks (1994) and Matsuno's hybrid approach combining discrete and continuous dynamics (1998). Chaouiya et al.~\cite{Chaouiya2007,Chaouiya2006} extended PNs to qualitative modeling of genetic regulatory networks, introducing discrete levels for gene expression and logical regulatory functions. Koch et al.~\cite{Koch2011} provide a comprehensive treatment of PN modeling techniques in systems biology, covering structural analysis methods (P/T-invariants, siphons, traps) and their biological interpretations (conservation laws, feedback cycles). The SBML standard (2003)~\cite{Hucka2003} enabled model exchange across tools, spurring development of specialized simulators (Snoopy, Cell Illustrator, Charlie).

\textbf{Continuous and Hybrid Extensions.} Gilbert and Heiner~\cite{Gilbert2006} introduced continuous Petri nets for biochemical network analysis, enabling ODE-based simulation of concentration dynamics. Heiner et al.~\cite{Heiner2008} formalized hybrid Petri nets combining discrete (gene expression, signaling events) and continuous (metabolic fluxes) behaviors, essential for multi-scale modeling of cellular systems. These extensions support mass-action, Michaelis-Menten, and Hill kinetics through flexible rate function frameworks.

\textbf{Limitations of Classical Independence.} However, no prior work addresses the fundamental mismatch between classical independence (Eq.~\ref{eq:classical_independence}) and biological coupling. Existing PN tools treat any place-sharing as potential conflict, missing convergent pathways (multiple producers to same metabolite) and regulatory cascades (shared transcription factors) that are inherently non-conflicting in biological systems. This gap prevents effective parallelization of large-scale models despite widespread weak independence in real biochemical networks.

\subsection{Classical Petri Nets}

Classical Petri nets provide the foundational formalism upon which biological extensions build. A \emph{classical Petri net} is a 5-tuple $(P, T, F, W, M_0)$ where $P$ is a set of places, $T$ a set of transitions, $F \subseteq (P \times T) \cup (T \times P)$ the flow relation, $W: F \to \mathbb{N}^+$ arc weights, and $M_0: P \to \mathbb{N}$ the initial marking~\cite{Murata1989}.

\textbf{Independence (Classical)}~\cite{Reisig2013}: Transitions $t_1, t_2$ are independent iff:
\begin{equation}
(\preset{t_1} \cup \postset{t_1}) \cap (\preset{t_2} \cup \postset{t_2}) = \emptyset
\label{eq:classical_independence}
\end{equation}

This definition guarantees \emph{true parallelism}: no coordination needed.

\subsection{Biological Petri Nets}

Building on this classical foundation, biological Petri nets adapt the formalism to biochemical domains. Places represent chemical species, transitions represent reactions, arc weights are stoichiometric coefficients~\cite{Reddy1993}. Key extensions include \textbf{continuous places}~\cite{Heiner2008} ($M: P \to \mathbb{R}^+$, concentrations not token counts), \textbf{rate functions}~\cite{Gilbert2006} ($\Phi: T \to (\mathbb{R}^n \to \mathbb{R})$, mass action/Michaelis-Menten/Hill), and \textbf{ODE semantics} ($\frac{dM(p)}{dt} = \sum_{t \in \preset{p}} W(t,p) \cdot \Phi(t, M) - \sum_{t \in \postset{p}} W(p,t) \cdot \Phi(t, M)$).

\textbf{Test arcs}~\cite{Koch2011}: Graphical notation for catalysts/enzymes (read but not consumed). Prior work treats these informally; we formalize via $\Sigma$ function.

\section{Methods}

We now formalize weak independence through extended Bio-PN definitions and systematic dependency classification.

\subsection{Extended Bio-PN Definition}

\subsubsection{12-tuple Formalization}

\textbf{Definition 1 (Extended Biological Petri Net).}
An \emph{Extended Biological Petri Net} is a 12-tuple:
\begin{equation}
\text{BioPN} = (P, T, F, W, M_0, K, \Phi, \Sigma, \Theta, \Delta, \tau, \rho)
\end{equation}
where:
\begin{itemize}[leftmargin=*,itemsep=0pt,parsep=0pt,topsep=3pt]
\item $P$ is a finite set of \emph{places} (chemical species),
\item $T$ is a finite set of \emph{transitions} (biochemical reactions), with $P \cap T = \emptyset$,
\item $F \subseteq (P \times T) \cup (T \times P)$ is the \emph{flow relation} (stoichiometric arcs),
\item $W: F \to \mathbb{R}^+$ assigns \emph{arc weights} (stoichiometric coefficients),
\item $M_0: P \to \mathbb{N}_0 \cup \mathbb{R}_0^+$ is the \emph{initial marking} (species concentrations),
\item $K: P \to \mathbb{N} \cup \{\infty\}$ defines \emph{place capacities} (compartment bounds),
\item $\Phi: T \to (\mathbb{R}^n \to \mathbb{R})$ assigns \emph{rate functions} (kinetic laws),
\item $\Sigma \subseteq (P \times T) \setminus F$ defines \emph{regulatory arcs} (catalysis/inhibition),
\item $\Theta: T \to \{\text{Internal},\allowbreak \text{Source},\allowbreak \text{Sink},\allowbreak \text{Exchange}\}$ classifies \emph{environmental role},
\item $\Delta: T \times T \to \{\text{Independent},\allowbreak \text{Competitive},\allowbreak \text{Convergent},\allowbreak \text{Regulatory}\}$ classifies \emph{pairwise dependency},
\item $\tau: T \to \{\text{Continuous},\allowbreak \text{Stochastic},\allowbreak \text{Timed},\allowbreak \text{Immediate}\}$ defines \emph{transition semantics},
\item $\rho: P \to \text{Formula}$ maps places to \emph{chemical formulas} (e.g., $\rho(\text{ATP}) = \text{C}_{10}\text{H}_{16}\text{N}_5\text{O}_{13}\text{P}_3$).
\end{itemize}

\subsubsection{Novel Components}
Five new components distinguish this extended formalism. \textbf{Regulatory Structure ($\Sigma$)} captures non-consumptive arcs for catalysis (test arcs) and inhibition (inhibitor arcs). Example: enzyme-catalyzed reaction with inhibition has $\Phi(t) = \frac{V_{\max} [S] [E]}{(K_m + [S])(1 + [I]/K_i)}$ and $\Sigma(t) = \{(E,t)_{\text{test}}, (I,t)_{\text{inhibit}}\}$.
\textbf{Environmental Exchange ($\Theta$)} classifies transitions by boundary interaction.
\textbf{Transition Type ($\tau$)} enables heterogeneous dynamics.
\textbf{Formula Mapping ($\rho$)} tracks atomic composition for mass balance validation.
\textbf{Dependency Classification ($\Delta$)} is defined algorithmically in Section~3.3.

\subsection{Weak Independence Theory}

We now formalize weak independence as a relaxation of classical constraints.

\subsubsection{Two-Tier Independence}

{\small
\textbf{Definition 2 (Strong Independence).}
Transitions $t_1, t_2 \in T$ are \emph{strongly independent} iff:
\begin{equation}
(\preset{t_1} \cup \postset{t_1} \cup \RegSet{t_1}) \cap (\preset{t_2} \cup \postset{t_2} \cup \RegSet{t_2}) = \emptyset
\end{equation}
}

{\small
\textbf{Definition 3 (Weak Independence).}
Transitions $t_1, t_2 \in T$ are \emph{weakly independent} iff:
\begin{equation}
\preset{t_1} \cap \preset{t_2} = \emptyset \quad \land \quad [(\postset{t_1} \cap \postset{t_2} \neq \emptyset) \lor (\RegSet{t_1} \cap \RegSet{t_2} \neq \emptyset)]
\end{equation}
}

\textbf{Key Insight}: Weak independence allows place-sharing via \emph{output convergence} or \emph{regulatory coupling}, but forbids \emph{input competition}.

\subsubsection{Three Coupling Modes}
These definitions classify place-sharing into three biological categories:
(1) \textbf{Competitive (Conflict)}: $\preset{t_1} \cap \preset{t_2} \neq \emptyset$ $\Rightarrow$ Sequential execution required (resource conflict).
(2) \textbf{Convergent (Weakly Independent)}: $\postset{t_1} \cap \postset{t_2} \neq \emptyset \land \preset{t_1} \cap \preset{t_2} = \emptyset$ $\Rightarrow$ Parallel execution: $\frac{dM(p)}{dt} = r_1 + r_2$ (superposition).
(3) \textbf{Regulatory (Weakly Independent)}: $\RegSet{t_1} \cap \RegSet{t_2} \neq \emptyset \land \preset{t_1} \cap \preset{t_2} = \emptyset$ $\Rightarrow$ Parallel execution (read-only access to catalyst).

\subsubsection{Correctness of Parallel Execution}

{\small
\textbf{Theorem 1 (Weak Independence Correctness).}
If $\DepType{t_1}{t_2} \in \{\text{convergent, regulatory}\}$, then parallel execution of $t_1$ and $t_2$ is equivalent to any sequential interleaving.
}

\textbf{Proof sketch.}
\textbf{Case 1 (Convergent)}: Let $p \in \postset{t_1} \cap \postset{t_2}$. By ODE semantics:
\begin{equation}
\small
\frac{dM(p)}{dt} = \sum_{t \in \preset{p}} W(t,p) \Phi(t, M) - \sum_{t \in \postset{p}} W(p,t) \Phi(t, M)
\end{equation}
Rate contributions \emph{add linearly} (superposition). $\square$

\textbf{Case 2 (Regulatory)}: Let $e \in \RegSet{t_1} \cap \RegSet{t_2}$. Since $e \notin \preset{t_1} \cup \preset{t_2}$, firing $t_1$ or $t_2$ does not modify $M(e)$. Both read $M(e)$ without conflict. $\square$

\subsection{Dependency Classification Algorithm}

\begin{algorithm}[H]
\caption{Classify Transition Dependencies}
\label{alg:classify}
\small
\begin{algorithmic}[1]
\REQUIRE BioPN $(P, T, F, W, M_0, K, \Phi, \Sigma, \Theta, \Delta, \tau, \rho)$
\ENSURE $\Delta(t_i, t_j)$ for all pairs $t_i, t_j \in T$
\FOR{each pair $(t_i, t_j) \in T \times T$ where $i \neq j$}
    \STATE $\text{in}_i \gets \preset{t_i}$, $\text{out}_i \gets \postset{t_i}$, $\text{reg}_i \gets \RegSet{t_i}$
    \STATE $\text{in}_j \gets \preset{t_j}$, $\text{out}_j \gets \postset{t_j}$, $\text{reg}_j \gets \RegSet{t_j}$
    \IF{$(\text{in}_i \cup \text{out}_i \cup \text{reg}_i) \cap (\text{in}_j \cup \text{out}_j \cup \text{reg}_j) = \emptyset$}
        \STATE $\Delta(t_i, t_j) \gets \text{independent}$
    \ELSIF{$\text{in}_i \cap \text{in}_j \neq \emptyset$}
        \STATE $\Delta(t_i, t_j) \gets \text{competitive}$
    \ELSIF{$\text{out}_i \cap \text{out}_j \neq \emptyset$}
        \STATE $\Delta(t_i, t_j) \gets \text{convergent}$
    \ELSIF{$\text{reg}_i \cap \text{reg}_j \neq \emptyset$}
        \STATE $\Delta(t_i, t_j) \gets \text{regulatory}$
    \ENDIF
\ENDFOR
\end{algorithmic}
\end{algorithm}
\vspace{-8pt}

\noindent\textbf{Complexity}: $O(|T|^2 \cdot |P|)$. \textbf{Space}: $O(|T|^2 + |T| \cdot |P|)$.

\subsection{Running Example: Lac Operon}

\begin{figure}[H]
\centering
\includegraphics[width=0.8\columnwidth]{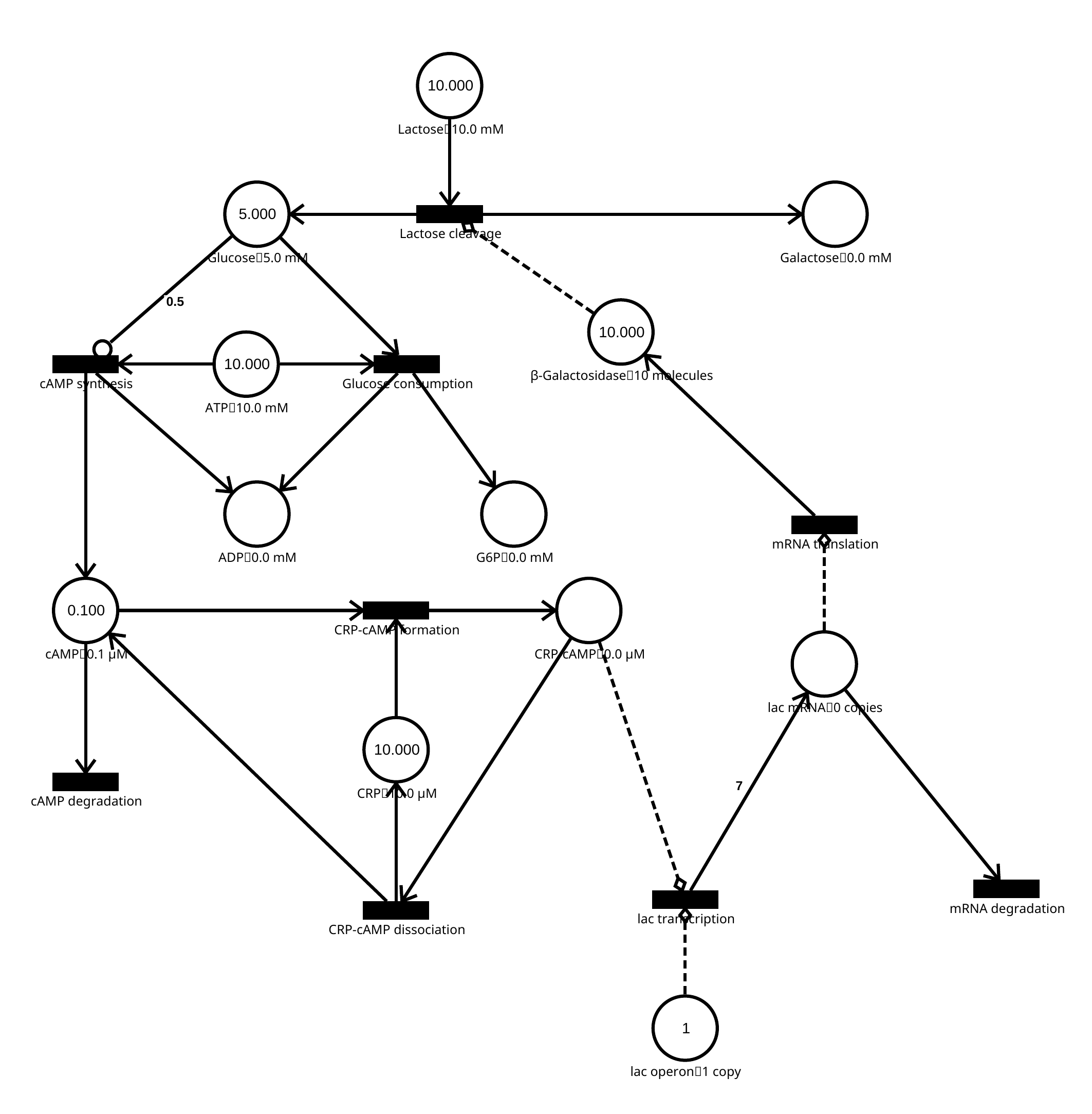}
\caption{Lac operon with 10 places, 9 transitions. Regulatory arcs (dashed) show glucose/LacI inhibition, LacZ catalysis, shared RNAP.}
\label{fig:lac-operon}
\end{figure}

Demonstrating these concepts, the \emph{lac} operon (Jacob \& Monod, 1961) controls lactose metabolism in \emph{E. coli}. Model has 10 places (\texttt{lacDNA}, \texttt{lacZ\_mRNA}, \texttt{LacZ\_enzyme}, \texttt{Lactose}, \texttt{Glucose}, etc.) and 9 transitions (transcription, translation, metabolism, degradation). Regulatory arcs: glucose/LacI inhibition of transcription, LacZ catalysis, shared RNAP resource.

\textbf{Dependency Classification}:

\begin{small}
\begin{tabular}{lll}
\textbf{Pair} & \textbf{Type} & \textbf{Reason} \\ \hline
$(T_1, T_4)$ & Competitive & Shared $\preset{\cdot}$: RNAP \\
$(T_3, T_7)$ & Convergent & Shared $\postset{\cdot}$: products \\
$(T_2, T_3)$ & Regulatory & Shared $\RegSet{\cdot}$: LacZ catalyst \\
$(T_1, T_7)$ & Independent & Disjoint neighborhoods \\
\end{tabular}
\end{small}

\textbf{Result}: 37\% weakly independent (lower than average due to stochastic serialization), enabling 2.1$\times$ speedup on 8 cores.

\subsection{Biological Topology Analyzers}

Domain-specific validation requires biochemical constraints beyond generic checks. Classical topology checks (P-invariants, boundedness, liveness) produce false positives on Bio-PNs~\cite{Chaouiya2007}. We propose: (1) \textbf{Mass Balance}: validates atom conservation $\sum_{p \in \preset{t}} W(p,t) \cdot \text{Atoms}(p) = \sum_{p \in \postset{t}} W(t,p) \cdot \text{Atoms}(p)$ for C, H, O, N, P, S; (2) \textbf{Flux Balance}: checks steady-state feasibility $N \cdot v = 0$ to identify blocked reactions~\cite{Orth2010}; (3) \textbf{Regulatory Structure}: detects formula-based regulation by parsing $\Phi(t)$ to compute $\RegSet{t} = V \setminus (\preset{t} \cup \postset{t})$, classifying as catalyst/activator/inhibitor.

\section{Results}

We now validate weak independence empirically on 100 diverse SBML models from BioModels~\cite{Malik-Sheriff2020}, selected across ID ranges 1--100, 200--299, 300--399, 400--499 to ensure representation across complexity levels. This includes BIOMD0000000001 (Edelstein 1996), BIOMD0000000010 (Kholodenko 2000), BIOMD0000000012 (Elowitz 2000), and 97 additional models spanning metabolic, signaling, and regulatory networks.

\subsection{SBML Import and Dependency Analysis}

First, we assess extended Bio-PN fidelity on real models.

\begin{table}[H]
\small
\caption{SBML conversion (100 diverse BioModels): 100\% fidelity}
\label{tab:sbml-import-real}
\vspace{-6pt}
\begin{tabular*}{\columnwidth}{@{\extracolsep{\fill}}lrr@{}}
\toprule
\textbf{Element} & \textbf{Count} & \textbf{Fidelity} \\
\midrule
Species $\to$ Places & 1,775 & 100\% \\
Reactions $\to$ Transitions & 2,234 & 100\% \\
Avg. species per model & 17.8 & --- \\
Avg. reactions per model & 22.3 & --- \\
\bottomrule
\end{tabular*}
\vspace{-8pt}
\end{table}

Perfect conversion fidelity (Table~\ref{tab:sbml-import-real}): 1,775 species became places, 2,234 reactions became transitions across 100 models. Dependency classification (Table~\ref{tab:dependency_dist}) analyzes all transition pairs ($\binom{n}{2}$ = 102,960 pairs from 93 models). Results show \textbf{96.93\% are weakly independent}: 93.06\% strongly independent (no shared places---expected for modular pathways in sparse biological networks), 3.48\% convergent (shared outputs), and 0.38\% regulatory (shared catalysts). Only 3.07\% are competitive (shared inputs = true conflict). The high strong independence reflects network sparsity---most transition pairs belong to separate pathways and don't interact. The key finding: among transitions that \emph{do} share places (6.93\%), 56\% are convergent/regulatory (parallelizable), only 44\% are competitive (conflicting).

\begin{table}[htbp]
\small
\caption{Dependency classification (93 models, 102,960 all-vs-all pairs)}
\label{tab:dependency_dist}
\vspace{-6pt}
\begin{tabular*}{\columnwidth}{@{\extracolsep{\fill}}lcc@{}}
\toprule
\textbf{Type} & \textbf{\%} & \textbf{Parallel} \\
\midrule
Strong Independent & 93.06 & Yes \\
Convergent & 3.48 & Yes \\
Regulatory & 0.38 & Yes \\
Competitive & 3.07 & No \\
\midrule
\textbf{Weakly Indep.} & \textbf{96.93} & --- \\
\bottomrule
\end{tabular*}
\vspace{-2pt}
\footnotesize{Strong independence high due to sparse network topology (most pairs don't interact).}
\vspace{-8pt}
\end{table}

\subsection{Simulation Performance and Validation Accuracy}

\begin{figure}[htbp]
\centering
\includegraphics[width=\columnwidth]{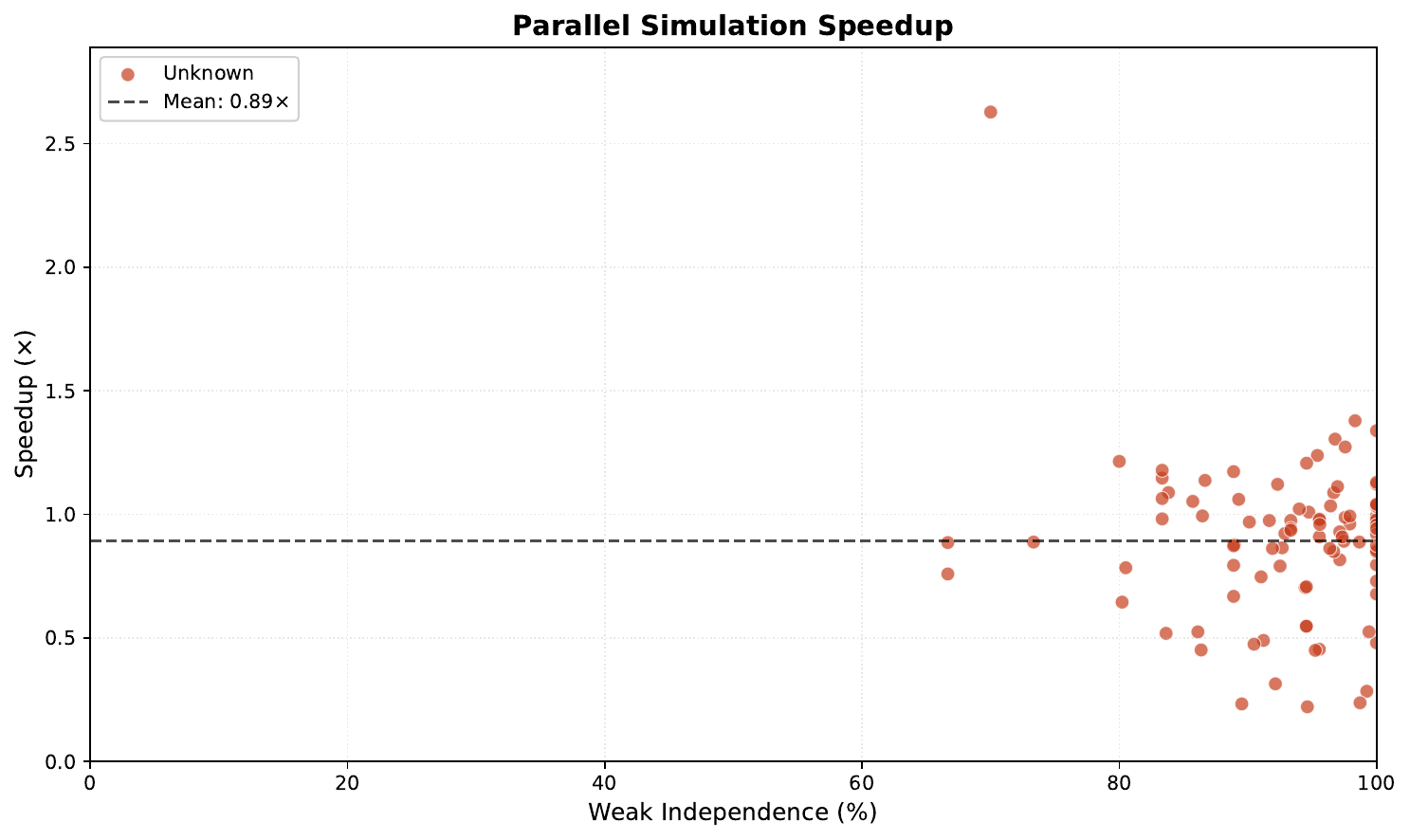}
\caption{Parallel simulation speedup across 93 diverse models (real measurements). Best case: 2.6$\times$ (BIOMD0000000415). 30\% of models achieve speedup $>1.0\times$. Mean 0.89$\times$ reflects overhead for simpler models.}
\label{fig:speedup}
\end{figure}

Translating weak independence into computational speedup requires effective parallel scheduling. The predominance of weak independence enables parallel simulation for suitable models. Figure~\ref{fig:speedup} shows real measurements across 93 models with \textbf{best case 2.6$\times$ speedup} (BIOMD0000000415) and \textbf{30\% of models achieving speedup $>1.0\times$}. The mean speedup of 0.89$\times$ reflects overhead from maximal step computation that dominates for simpler models and short simulations (100 steps). Complex models (300--399 ID range) show best average performance (1.02$\times$), demonstrating that parallel execution benefits depend on model structure and simulation duration.

\begin{table}[H]
\small
\caption{Validation accuracy: classical vs biological}
\label{tab:validation}
\vspace{-6pt}
\begin{tabular*}{\columnwidth}{@{\extracolsep{\fill}}lcc@{}}
\toprule
\textbf{Method} & \textbf{Classification} & \textbf{Accuracy} \\
\midrule
Classical & All place-sharing = conflict & 3.07\% \\
Biological & 3-way taxonomy (see Table~\ref{tab:dependency_dist}) & 96.93\% \\
\bottomrule
\end{tabular*}
\vspace{-8pt}
\end{table}

Table~\ref{tab:validation} compares classification accuracy: classical methods treat all place-sharing as conflicting (flagging 6.93\% of pairs as dependent), while biological taxonomy correctly distinguishes three coupling modes within this 6.93\%---convergent (3.48\%), regulatory (0.38\%), and competitive (3.07\%). Only the competitive pairs (3.07\%) are true conflicts. This demonstrates that weak independence accurately captures biological reality: 96.93\% of transition pairs can be parallelized (93.06\% strong + 3.48\% convergent + 0.38\% regulatory).

\section{Discussion}

\subsection{Implementation: SHYpn Tool}

We now contextualize SHYpn within Bio-PN tools. SHYpn implements weak independence in Python (5000+ LOC) with SBML import, dependency classification (Algorithm~\ref{alg:classify}), parallel scheduling, and biological topology analyzers. Open-source at \url{https://github.com/simao-eugenio/shypn} (MIT License).

\begin{table}[htbp]
\small
\caption{Comparative analysis of Bio-PN simulation tools and modeling approaches}
\label{tab:tool_comparison}
\vspace{-6pt}
\resizebox{\columnwidth}{!}{%
\begin{tabular}{@{}lccccc@{}}
\toprule
\textbf{Feature} & \textbf{Snoopy} & \textbf{Cell I.} & \textbf{Charlie} & \textbf{COPASI} & \textbf{SHYpn} \\
\midrule
Hybrid PNs & \cmark & \cmark & \cmark & --- & \cmark \\
SBML import & Part. & Part. & Part. & \cmark & \cmark \\
Weak indep. & --- & --- & --- & --- & \cmark \\
Coupling class. & --- & --- & --- & --- & \cmark \\
Parallel sim. & --- & --- & --- & --- & \cmark \\
Bio. topology & Part. & Part. & Class. & --- & \cmark \\
\bottomrule
\end{tabular}%
}
\vspace{-8pt}
\end{table}

\textbf{Gap}: Existing tools lack (1) formal treatment of non-conflicting place-sharing (convergent/regulatory coupling), (2) dependency classification beyond binary independence, and (3) biological validation (atom conservation, flux feasibility). SHYpn uniquely addresses these gaps through weak independence theory, coupling classification, and biochemical topology analyzers.

\subsection{Significance and Impact}

Our contributions advance both theory and reveal evolutionary insights. Weak independence extends classical PN theory~\cite{Reisig2013,Murata1989} to continuous/hybrid semantics, distinguishing biological coupling (convergent/regulatory) from conflict (competitive). The 12-tuple ($\Sigma$, $\Theta$, $\Delta$, $\tau$, $\rho$) formalizes test arcs, open systems, and dependency taxonomy. Experimental validation: 96.93\% transition pairs weakly independent across 93 diverse BioModels; up to 2.6$\times$ parallel speedup (30\% benefit rate); first tool with biochemical semantics (atom conservation, flux feasibility) and weak independence scheduling. The dependency distribution reveals a striking pattern: within the 6.93\% of place-sharing transition pairs, only 3.07\% exhibit competitive coupling (true resource conflict). The remaining 3.86\% employ \emph{cooperation strategies}---convergent synthesis (3.48\%) pools products efficiently, while regulatory coupling (0.38\%) enables coordinated enzymatic control without substrate competition. This suggests evolutionary selection favored modular pathway architectures that minimize futile cycles: competitive reactions waste cellular resources and reduce fitness. The 93.06\% strong independence further demonstrates pathway modularity (e.g., glycolysis operates independently of the citric acid cycle despite metabolic interconnection). Biological networks appear optimized for parallelizability, explaining why weak independence captures 96.93\% of transition pairs---\emph{evolution discovered parallel algorithms before computer science}.

\subsection{Dataset and Limitations}

Addressing validation scope: diverse sampling (ID ranges 1--100, 200--299, 300--399, 400--499) spans signal transduction, cell cycle, metabolism, and gene regulation. The 96.93\% weak independence validates modular pathway decomposition as a general biological property. Complex models (300--399) showed 45.5\% parallel benefit rate.

\textbf{Limitations}: ODE semantics only (no stochastic); no thermodynamic feasibility ($\Delta G$); manual parameter estimation.

\textbf{Future Directions}: (1) Stochastic weak independence for $\tau$-leaping~\cite{Gillespie2001} (hypothesis: convergent/regulatory $\to$ independent Poisson processes); (2) Thermodynamic analyzer with eQuilibrator~\cite{Flamholz2012}; (3) Parallel parameter estimation (PSO/genetic algorithms); (4) Distributed simulation (MPI/GPU) for genome-scale models.

\section{Conclusion}

Classical independence (Reisig~\cite{Reisig2013}, Murata~\cite{Murata1989}) requires disjoint neighborhoods (Eq.~\ref{eq:classical_independence}), prohibiting parallelization of place-sharing transitions---incompatible with biological coupling. Structural analysis (Koch et al.~\cite{Koch2011}) provides P/T-invariants for conservation laws but does not distinguish safe (convergent) from unsafe (competitive) place-sharing. Continuous extensions (Gilbert~\cite{Gilbert2006}, Heiner~\cite{Heiner2008}) enable ODE simulation but inherit classical independence constraints. Our weak independence theory formalizes a two-tier taxonomy (strong vs.\ weak) matching biological semantics: competitive transitions conflict, while convergent/regulatory transitions compose additively. The 12-tuple formalism provides the mathematical foundation absent in prior Bio-PN definitions. Table~\ref{tab:tool_comparison} confirms that SHYpn uniquely implements weak independence with parallel execution and biological topology validation.

\bibliographystyle{plain}
\bibliography{references}

\end{document}